\documentclass{iau}
\usepackage{graphicx}
\title[Galaxy alignment] 
{Galaxy alignment on large and small scales}

\author[X. Kang et al.]  
{X. Kang$^{1}$\thanks{Email: {\tt kangxi@pmo.ac.cn} },
W.P. Lin$^{2}$,
X. Dong$^{2}$,
Y.O. Wang$^{2}$,
A. Dutton$^{3}$ 
and A. Macci\`o$^{3}$
}
\affiliation{
$^{1}$Purple Mountain Observatory, the Partner Group of MPI f\"{u}r Astronomie,  2 West Beijing Road, Nanjing 210008, China \\
$^{2}$Key Laboratory for Research in Galaxies and Cosmology, Shanghai Astronomical Observatory, Chinese Academy of Science, 80 Nandan Road, Shanghai 200030, China\\
$^{3}$Max Planck Institut f\"{u}r Astronomie, K\"{o}nigstuhl 17, D-69117 Heidelberg, Germany\\
}

\pubyear{2014}
\volume{308}  
\pagerange{119--126}
\jname{The Zeldovich Universe,
Genesis and Growth of the Cosmic Web}
\editors{Rien van de Weygaert, Sergei Shandarin, Enn Saar \& Jaan Einasto}
\begin{document}

\maketitle

\begin{abstract}
Galaxies are not randomly distributed across the universe but showing different kinds of alignment on different scales. On small scales satellite galaxies have a tendency to distribute along the major axis of the central galaxy, with dependence on galaxy properties that both red satellites and centrals have stronger alignment than their blue counterparts. On large scales, it is found that the major axes of Luminous Red Galaxies (LRGs) have correlation up to 30Mpc/h. Using hydro-dynamical simulation with star formation, we investigate the origin of galaxy alignment on different scales. It is found that most red satellite galaxies stay in the inner region of dark matter halo inside which the shape of central galaxy is well aligned with the dark matter distribution. Red centrals have stronger alignment than blue ones as they live in massive haloes and the central galaxy-halo alignment increases with halo mass. On large scales, the alignment of LRGs is also from the galaxy-halo shape correlation, but with some extent of mis-alignment. The massive haloes have stronger alignment than haloes in filament which connect massive haloes. This is contrary to the naive expectation that cosmic filament is the cause of halo alignment.


\end{abstract}

\firstsection 
\section{Introduction}

The spatial and orientation of galaxies are not randomly distributed in the universe. On small scales the distribution of satellite galaxies is correlated with the central galaxy.  For example, in our Milky Way Galaxy and neighboring M31, the satellites are found to be highly anisotropic that most of them are distributed in a great thin plane with a common rotation (e.g., Kroupa et al. 2005;  Ibata et al. 2013). This raises  questions  if our Milky Way or M31 is anomaly or the formation and accretion of its satellites is along some special direction, or even the Cold Dark Matter theory is wrong (e.g., Kang et al. 2005; Pawlowski et al. 2014).  It is highly possible that the Milky Way or M31 could be an outlier, thus its unique may not violate the general predictions from CDM model. However, the large sky surveys, such as 2dFGRS and SDSS, have found that satellites are still not randomly distributed around central galaxies, but are preferentially distributed along the major axes of centrals. This phenomenon is know as galaxy alignment. (e.g., Yang et al. 2006).  On large scales, it is found that the shape of Luminous Red Galaxies (LRGs) are correlated  up to scales of 30$\sim$70Mpc/h (Okumura et al. 2009; Li et al. 2013).

Most studies using N-body simulations are devoted to the origin of galaxy alignment, and they have found that the observed alignment  can be reproduced if the shape of central galaxies are correlated with that of the dark matter halo (e.g., Kang et al. 2007).  However, to match the color dependence of galaxy alignment, one has to make different assumption of how blue and red centrals follow the dark matter haloes (Agustsson \& Brainerd 2010). Thus no consistent conclusions about the color dependence of galaxy alignment have been drawn. To study the galaxy alignment in detail and avoid making assumption about how the shape of central galaxy follow the dark matter halo in N-body simulation, we make use of hydro-dynamical simulation which includes important physics governing galaxy formation, such as gas cooling, metal recycling, star formation and supernova feedback. As the simulation includes star formation, the shape of central galaxy and the properties of satellites are self-consistently included in the simulation, thus the predictions can be directly tested against the data.

\section{Simulation and Methods}

The hydro-dynamical simulation we use is run using the Gadget-2 code (Springel 2005), and it follows $2 \times 512^{3}$ dark matter particles and gas particles in a cube box with each side of $100Mpc/h$. The cosmological parameters are selected as $\Omega_{m}=0.268$, $\Omega_{\Lambda}=0.732$, $\sigma_{8}=0.85$ and $h=0.71$.   The dark matter haloes and galaxies are both found using the standard friends-of-friends (FoF) algorithm. In each FoF group, the most massive galaxy is defined as the central galaxy, and all others are defined as satellites. The reduced inertia tensor of  central galaxy determines its shape and major axis.  The distribution of satellites around central is described by the angle between the position of satellite galaxy respect to the major axis of central galaxy. An alignment is obtained if the average angle is less than $45^{\circ}$. For more details, see Dong et al. (2014).

\section{Results}

\begin{figure}[h]
\begin{center}
\includegraphics[width=0.8\textwidth, height=0.4\textwidth]{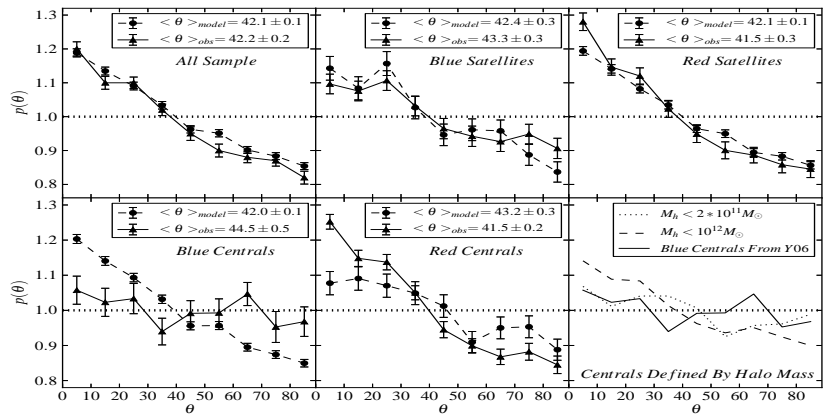}
\caption{Predicted galaxy alignment (circles connected by dashed lines). The solid lines with triangles show the observational results of Yang et al. (2006) from SDSS. Each panel shows different samples.}
\end{center}
\end{figure}

Fig.1 shows the predicted alignment with comparison to the data of Yang et al. (2006).  The upper left panel shows that the predicted alignment agrees well with the data for all galaxies. Also the simulation reproduces the dependence on color of satellites, as shown in upper middle and right panels that red satellites have stronger alignment. The lower left and middle panels show the alignment for blue and red centrals, and it is found that the predictions are inconsistent with the data. The predicted signal for blue centrals is too high and the one for red centrals is too low. We found that this is due to the neglect of effective feedback in our simulation, such as AGN feedback, which results in too many massive blue central galaxies (see Fig.1 in Dong et al. 2014). Those blue centrals are living in massive haloes where the alignment is larger (see right panel of Fig.2).

It is known observationally that the color of central galaxy is strongly correlated with the host halo mass (e.g., Yang et al. 2008), and our simulation fails to reproduce that correlation.  To mimic this effect and see if the dependence on color of central is from the halo mass dependence, in the lower right panel of Fig.1 we show the prediction for centrals in host halo mass lower than some critical mass (see caption in plot). It is found that the alignment of centrals is a function of halo mass, and if blue centrals are living in haloes with mass lower than $2\times 10^{11}M_{\odot}$, the predicted alignment is close to the data. Such a critical mass is also obtained from hydro-dynamical simulations (e.g., Kere{\v s} et al. 2005).

Fig.2 explains the origin of alignment on galaxy color. The right panel  shows the alignment between the shape of central galaxy and  dark matter halo inside different radii. It is found that central galaxy traces better the shape of dark matter in inner region (red dotted line) and this alignment increases with halo mass. The left panel plots the radial distribution of red/blue satellites (with red solid and blue dashed lines). It is found that most red satellites stay in the inner halo region and thus trace the shape of dark matter halo there. As central galaxy follows better the shape of halo in inner region, it naturally predicts that red satellites are stronger aligned with central than blue satellites. The middle panel shows the alignment for satellites at different distance to central galaxy and good agreement with data is seen. The halo mass dependence seen in the right panel also explains the observed dependence on color of centrals as most red centrals live in massive haloes. 

\begin{figure}[t]
\begin{center}
\includegraphics[width=0.9\textwidth]{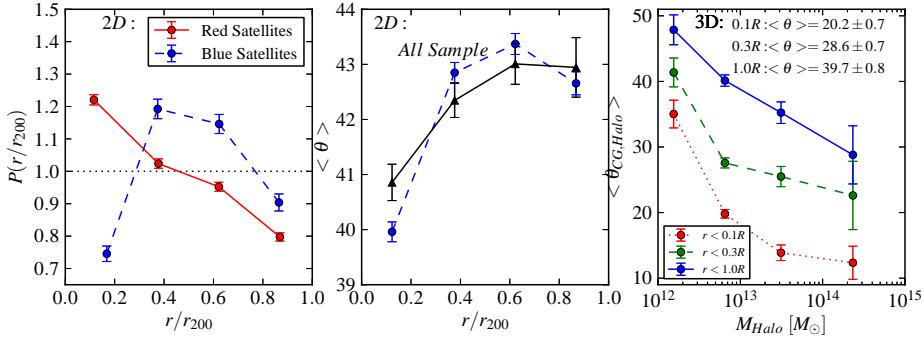}
\caption{Left panel: satellites radial distribution. Middle: alignment of satellite in different radii. Right panel: central galaxy-halo shape alignment, with color lines for halo shape measured inside different radii.}
\end{center}
\vspace {-0.4 cm}
\end{figure}

On large scales, the galaxy alignment is often referred to the shape correlation of central galaxies. It is found that the major axes of LRGs at z$\sim 0.3$ are correlated up to 30 $\sim $70Mpc/h (Okumura et al. 2009; Li et al. 2013), as shown by the points in upper left panel of Fig.3. Okumura et al. also found that if the major axis of LRG follows that of the dark matter perfectly, the predicted alignment from N-body simulation is higher than the data (red dashed line), but if there is an average misalignment about $35^{\circ}$, the prediction will match well with the data (red solid line). As the LRGs often live in massive haloes connect by cosmic filament, it is natural to ask if the halo shape correlation is related to haloes in filament.The right panel of Fig.3 shows that shape correlation of haloes in different environment (Zhang et al. 2008). It is found that haloes in dense region have stronger correlation than haloes in filament.  Thus the observed alignment of LRGs seems to be fixed before the formation of the filaments connecting them.

\begin{figure}[t]
\includegraphics[width=0.4\textwidth, height=0.4\textwidth]{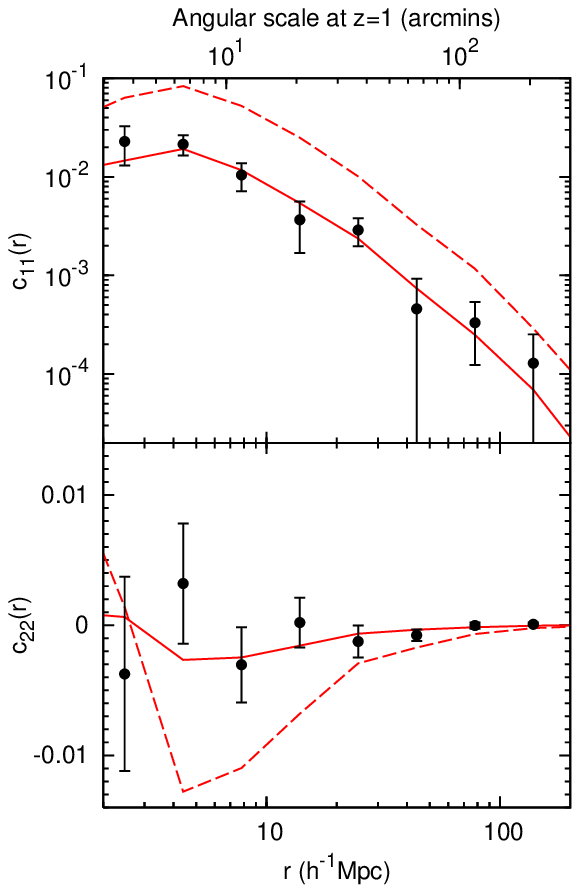}
\hfill
\includegraphics[width=0.4\textwidth, height=0.4\textwidth]{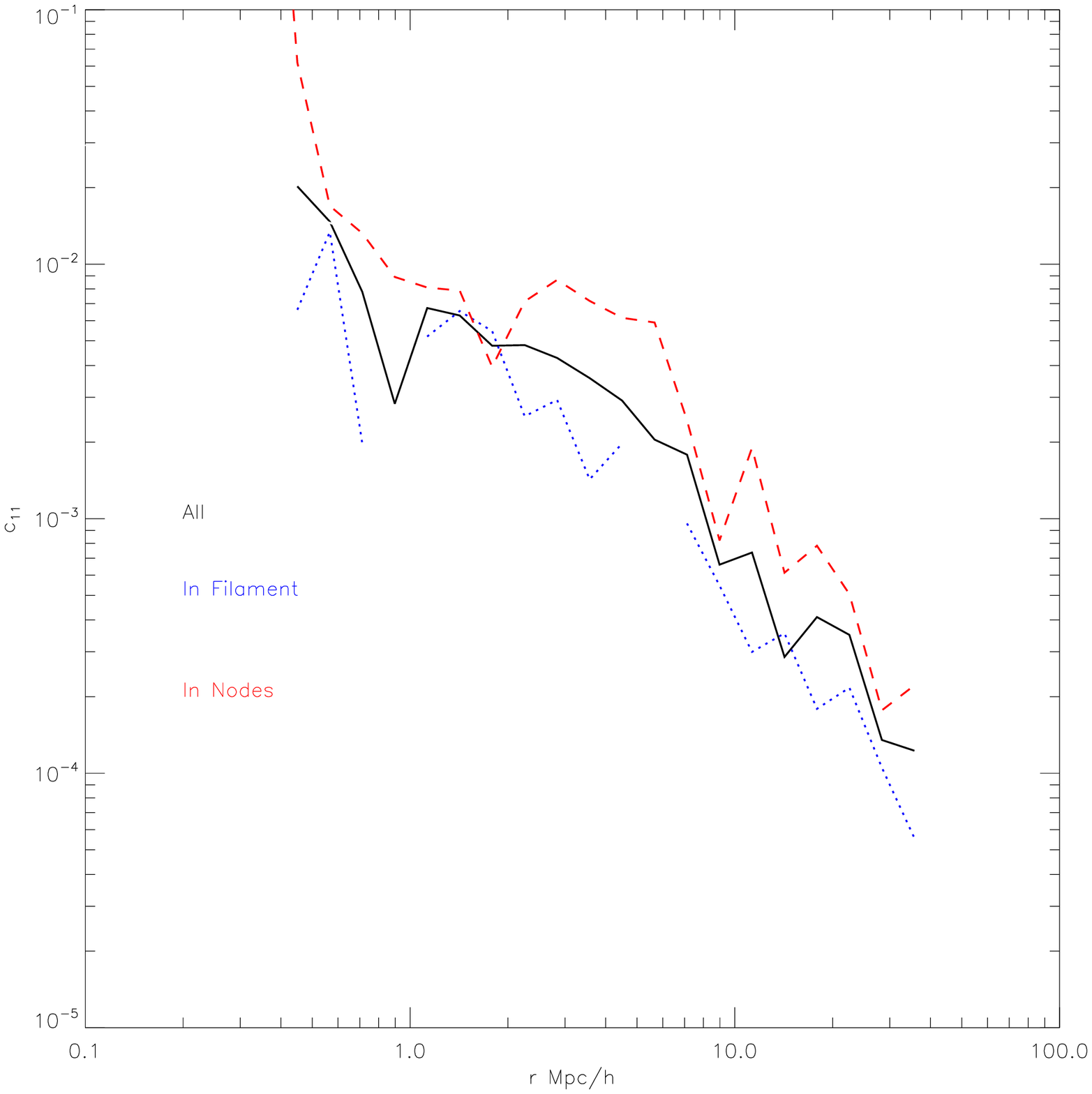}
\caption{Left panel: the shape correlation of LRGs. Upper panel shows that LRGs are correlated along their major axes, but not along minor axes (lower panel). Right panel: halo shape correlation in different environment. Stronger alignment is found for haloes in dense environment.}
\end{figure}

\section{Conclusion}
Galaxy alignment effects are seen on both large and small scales. On small scales, it is mainly caused by the non-spherical nature of dark matter formed in CDM scenario (Jing \& Suto 2002). The stronger alignment of red satellites is because they prefer halo inner region inside which the shape of central galaxy and dark matter distribution is well aligned, thus leading to stronger alignment between red satellites and central galaxy. The dependence on color of central is from the combined effects that red centrals live in massive haloes and the alignment between central galaxy and halo shape increases with halo mass.   On large scales, the shape correlation of LRGs is determined by the halo-halo shape correlation, but with an average misalignment about $35^{\circ}$. The shape correlation of haloes in dense region is larger than those in filament which connect the massive haloes.
 
 \vskip 0.5cm
{\sl  This work is partially supported by the NSFC project No.11333008 and the Strategic Priority Research Program "The Emergence of Cosmological Structures" of the Chinese Academy of Science Grant No. XDB09000000.}

\end{document}